# Indentation of concave power law profiles with arbitrary exponents

Valentin L. Popov, Markus Heß, Emanuel Willert, and Qiang Li

Technische Universität Berlin, Str. des 17. Juni 135, 10623 Berlin, Germany

**Abstract.** We study analytically and numerically the process of indentation of cylindrical rigid indenter with concave face in form of a power-law function. In the well-known case of a parabolic concave indenter, the contact starts at sharp edges of the indenter and spreads inwards with increasing indentation depth. For all profiles with the exponent larger than 2, the contact area first spreads from the boarder inwards, but then a contact is established in the center of the indenter. Finally, the outer ring spreads inwards and the central contact area outwards until the complete contact is achieved. The critical indentation depth for the full contact is calculated ones proceeding from the full contact and looking for the condition of vanishing pressure and also proceeding from incomplete contact (in this case numerically, using Boundary Element Method). The results of both approaches coincide.

## 1. Introduction

We consider indentation of a rigid cylindrical stamp with radius $a$ having concave face as shown in Fig. 1 with an elastic half-space. When a concave indenter is first brought into contact with the half-space, the initial contact has the shape of a thin ring of radius $a$. With increasing indentation, the outer boarder of the contact area remains the circle with radius $a$, while the inner border propagates inwards. The character of further changing the contact area depends on the detailed shape of the concave "face" of the indenter. In the case of a parabolic concave profile, the contact area spreads inwards continuously and the last point which comes into contact is in the center of indenter. However, for other shapes of concave profiles this is not necessarily the case. In particular, this is not the case for all power-law profiles (definitions of notations see in Fig. 1.)

$$\tilde{z} = f(r) = \begin{cases} -\dfrac{hr^n}{a^n}, & r \leq a, \\ \infty, & r > a \end{cases} \tag{1}$$

In the next sections, we first recapitulate and discuss the known solution for parabolic concave profile and then consider the cases of power-law profiles with exponents $n > 2$ and with $1 < n < 2$.

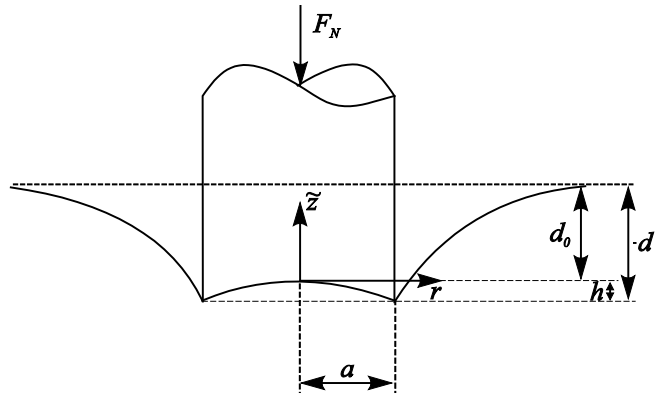

**Fig. 1 Normal indentation by a parabolic concave indenter.**

## 2. Parabolic concave profile (n=2)

Consider parabolic concave profile in the form

$$f(r) = \begin{cases} -\dfrac{hr^2}{a^2}, & r \leq a, \\ \infty, & r > a \end{cases} \qquad (2)$$

We calculated the indentation of this profile using the Boundary Element Method as described in [1]. In Fig. 2a, a series of contact area shapes is shown corresponding to five values of normalized "contact length" $L$ (definition see in [2].) The value "1" of the normalized contact length corresponds to the complete contact. One can see that the contact area has initially the shape of a thin ring which inner boundary propagates inwards with increasing indentation depth.

Fig. 2b shows the corresponding stress distributions: non-zero pressure in the contact ring with a square-root singularity at the outer boundary and vanishing pressure inside the ring. When the indentation depth increases, the region of zero pressure shrinks. In the critical state, when the first complete contact is achieved for the first time, the pressure is zero only in one single point $r = 0$ in the center of the contact (curve 5 in Fig. 2b). Thus, the critical state can be defined as the state of the complete contact with vanishing pressure in the center of the contact area.

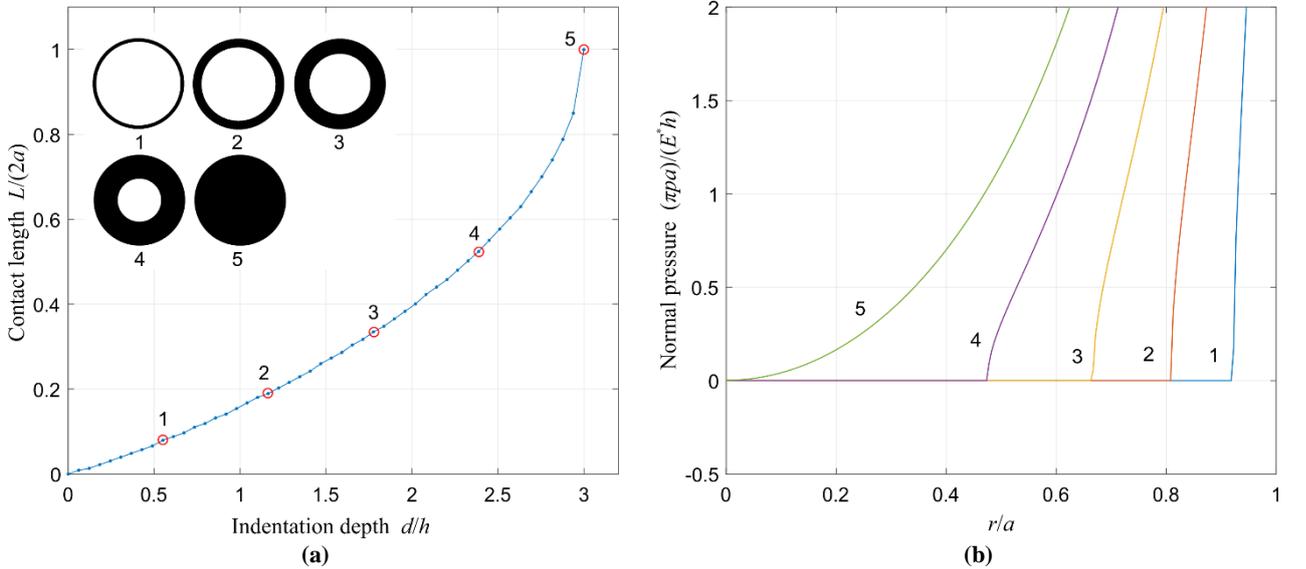

**Fig. 2** (a) Change in contact area during the indentation test of a parabolic concave profile; (b) the pressure distributions at 5 different indentation depth as marked with circles in (a).

One can obtain the criterion for the critical condition for the complete contact also by approaching the critical state from large indentation depths. In the state of a complete contact, the pressure distribution can be determined analytically. This solution was first found by Schubert in 1942 [3] (see also Barber, [4]). The simplest way is to use the Method of Dimensionality Reduction [5]. The explicit solution for pressure is given in [5]:

$$\bar{p} := \frac{\pi p a}{E^* h} = \frac{\delta + 1}{\sqrt{1 - \rho^2}} - 4\sqrt{1 - \rho^2}, \quad \text{with} \quad \delta := \frac{d}{h}, \; \rho = \frac{r}{a}. \qquad (3)$$

This dependency is presented in Fig. 3 for three values of normalized indentation depth. For $d/h > 3$ (not shown) the pressure is positive in all contact points. At $d/h = 3$ it becomes zero in the center of the contact. For smaller indentation depth the pressure becomes negative which indicate that the contact gets lost.

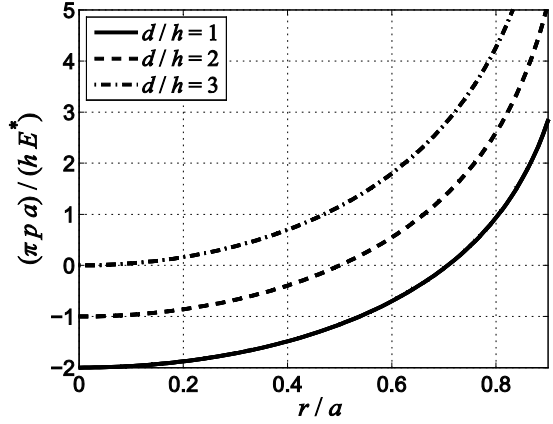

**Fig. 3** Pressure distribution under parabolic concave indenter for three various indentation depths (complete contact).

## 3. Power-law concave profiles with n>2

For power-law profiles with the exponent $n > 2$, the contact in the center of the indenter is achieved earlier than the complete contact. We illustrate this on the example of a concave power-law profile with $n = 4$. In this case, the pressure distribution is given by

$$\overline{p} = \frac{3\delta + 5}{3\sqrt{1-\rho^2}} - \frac{32}{9}\sqrt{1-\rho^2}\left(1 + 2\rho^2\right). \tag{4}$$

This pressure distribution is shown in Fig. 4. In this case, the minimum of pressure is not in the center of the contact so that if we start with large indentation depth and let it decrease, the pressure becomes zero not in the center but at finite radius $r/a = 1/2$. This occurs first when $d/h = 7/3 \approx 2.333$.

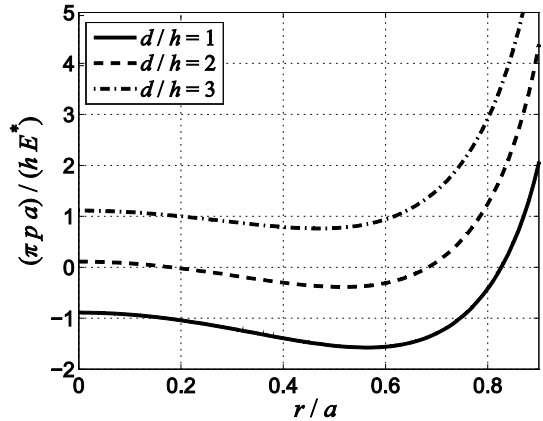

**Fig. 4** Pressure distribution for n = 4 (complete contact).

The sequence of contact configurations for $n = 4$ is shown in Fig. 5a. Initially, the contact area is a thin ring which is then expanding inwards. However, at some indentation depth a contact is established in the center of the indenter (point 3 in Fig. 5a). After that point, the central contact area expands outwards and the outer contact area propagates inwards until they merge.

Fig. 5b presents the pressure distributions corresponding to various in indentation depths. Observing the curve of pressure distribution, from the point 3 the pressure at the center is not zero any more. Due to the contact a "bump" appears in the center.

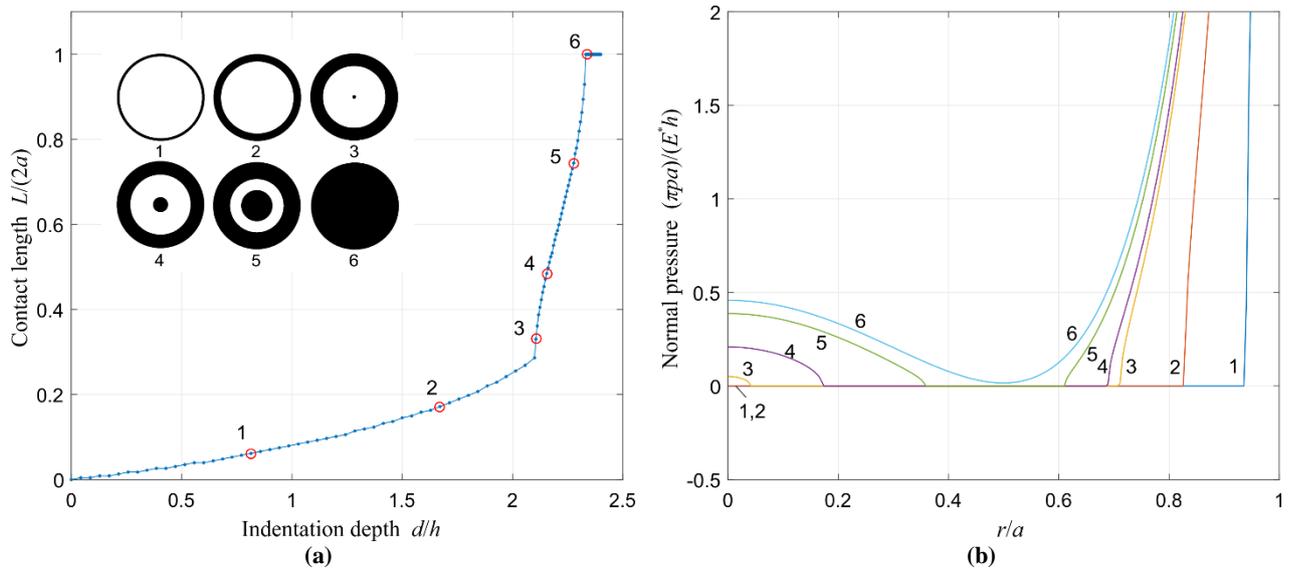

**Fig. 5 (a) Change in contact area during the indentation test of a power-law concave profile with exponent n=4; (b) the pressure distributions at 6 different indentation depth as marked with circles in (a).**

Using the MDR solution for arbitrary $n$ from [5], one can easily calculate the critical indentation depth for arbitrary (not necessarily integer) $n$. The results are presented in Fig. 6 and Tab. 1.

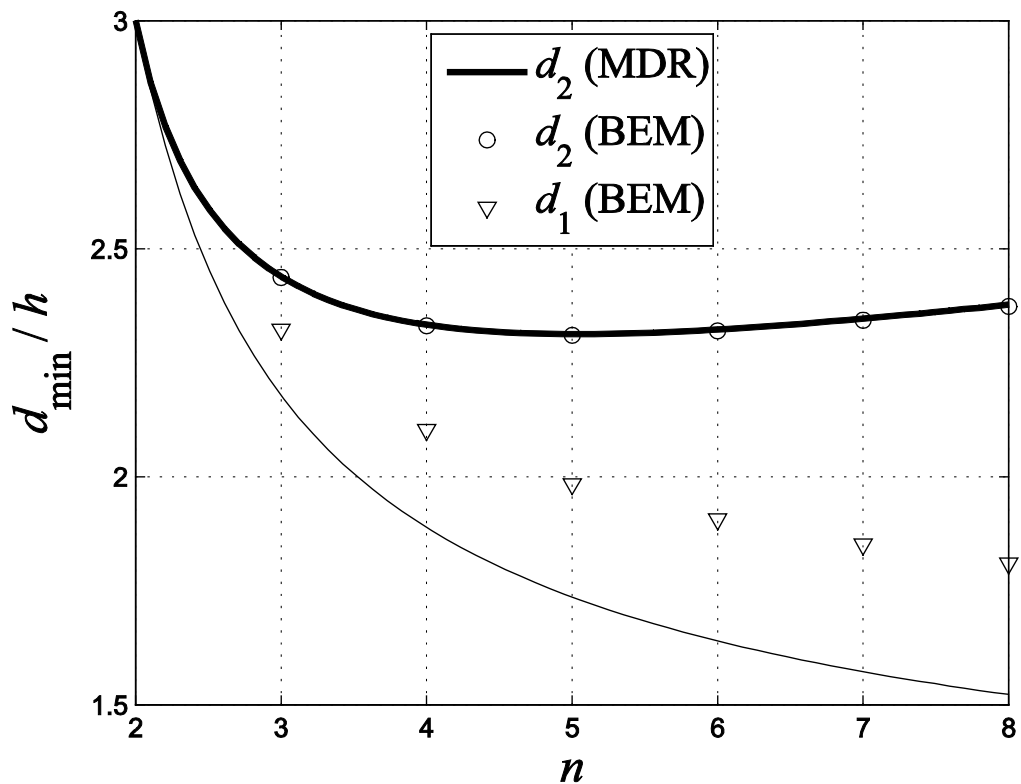

**Fig. 6 Critical indentation depths depending on $n$; $d_1$ is the indentation depth at which the first contact in the center of the indenter is established. This critical depth was calculated using BEM. $d_2$ is the depth at which the complete contact is achieved. It was calculated both using MDR and BEM. The solid line shows the theoretical estimation of the critical depth under assumption that it corresponds to zero pressure at point $r=0$ in a complete contact. This solution given in [5] is regrettably incorrect.**

**Tab. 1 Critical values of t $d_1/h$ and $d_2/h$ for $n>2$ according to BEM simulations**

| n | 2 | 3 | 4 | 5 | 6 | 7 | 8 |
|---|---|---|---|---|---|---|---|
| $d_1/h$ | 3 | 2.323 | 2.103 | 1.984 | 1.907 | 1.852 | 1.810 |
| $d_2/h$ | 3 | 2.437 | 2.333 | 2.310 | 2.320 | 2.343 | 2.373 |

## 4. Power-law concave profiles with 1<n<2

As in the case $n=2$, for all $1<n<2$ the contact starts with narrow ring at $r=a$ and spreads inward until the last contact is established in the center of the indenter. In all these cases the solution provided in [5] is valid and the critical depth for achieving complete contact is given by

$$d > d_{c,n} = h\left(1 + \frac{\kappa(n)}{n-1}\right) \tag{5}$$

with

$$\kappa(n) := \sqrt{\pi}\frac{\Gamma(n/2+1)}{\Gamma[(n+1)/2]}, \tag{6}$$

with the gamma function $\Gamma(\bullet)$

$$\Gamma(z) := \int_0^\infty t^{z-1}\exp(-t)\,dt. \tag{7}$$

Fig. 7.provides an illustration for the indentation of a concave power-law profile with $n=1.5$.

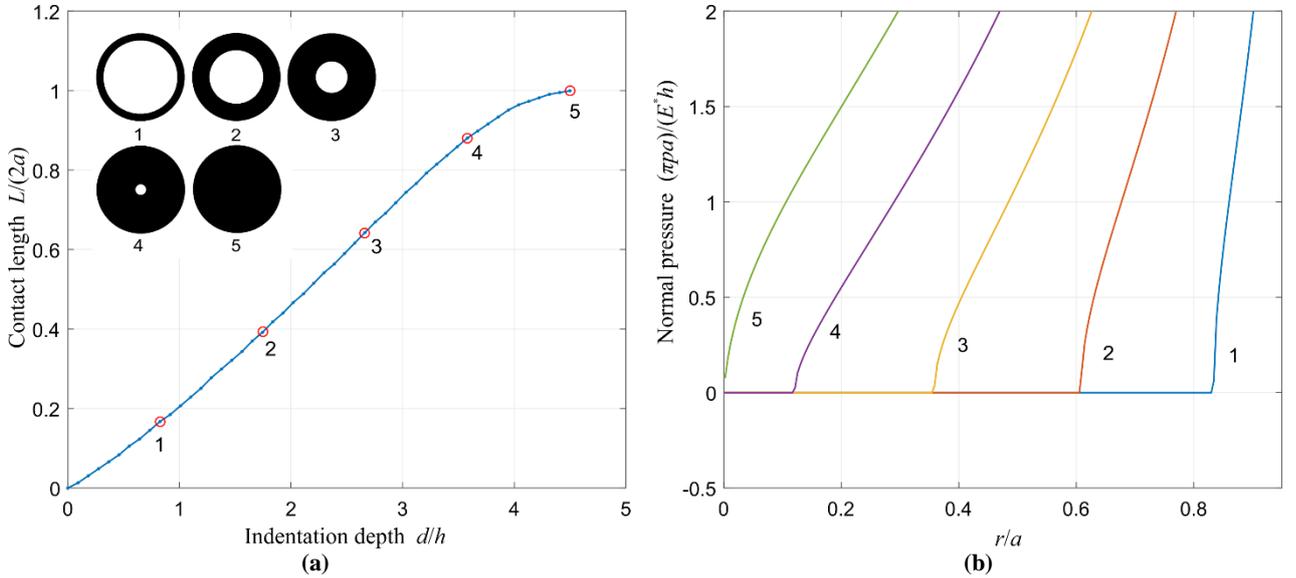

**Fig. 7** (a) Change in contact area during the indentation test of a power-law concave profile with the exponent *n*=1.5; (b) the pressure distributions for 5 different indentation depths as marked with circles in (a).

For large indentation depth (complete contact), the pressure distribution for n = 1.5 can be found in the analytical form:

$$\bar{p} = \frac{\delta - 1 + \kappa(1.5)}{\sqrt{1-\rho^2}} - 3\kappa(1.5)\left[ {}_2F_1\left(\frac{1}{2}, -\frac{1}{4}; \frac{3}{4}; \rho^2\right) - \frac{\kappa(1.5)}{3}\sqrt{\rho} \right],$$

$$\kappa(1.5) = \sqrt{\pi}\frac{\Gamma(7/4)}{\Gamma(5/4)}.$$

(8)

where

$$ {}_2F_1(a,b;c;z) := \sum_{n=0}^{\infty} \frac{(a)_n (b)_n}{(c)_n} \frac{z^n}{n!}, \quad |z| < 1 $$

(9)

is the hypergeometric function with Pochhammer symbol

$$(x)_n := \frac{\Gamma(x+n)}{\Gamma(x)}.$$

(10)

The pressure distribution is plotted in Fig. 8. One can see that in this case the minimum pressure is clearly in the center of the contact, and the contact gets lost when this pressure vanishes.

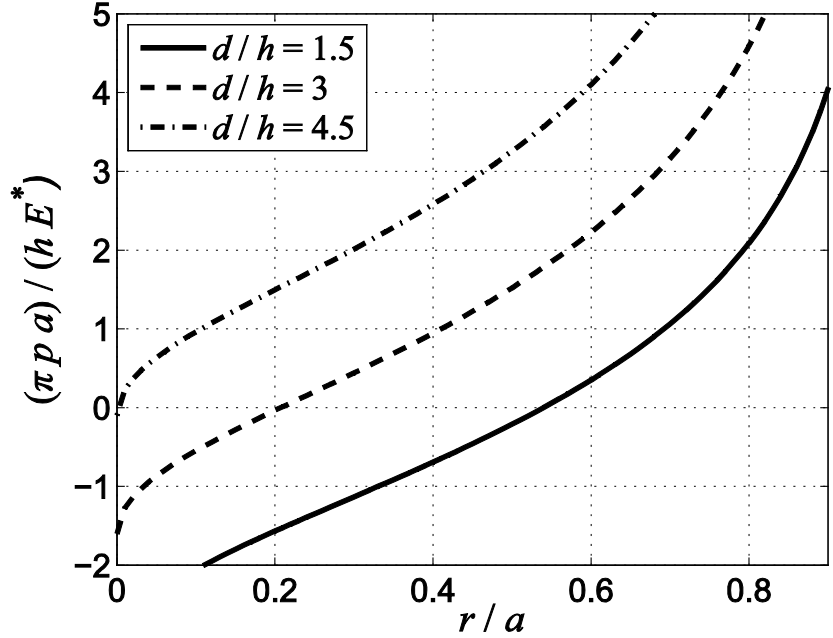

**Fig. 8** Pressure distribution for n = 1.5 and various indentation depths (complete contact)

## 5. Power-law concave profiles with n<1

As follows from (5), for $n < 1$ complete contact cannot be achieved independently on the indentation depth.

## 6. Conclusion

We provided analytical and numerical solution for the problem of indentation of a power-law concave profile and found the critical indentation depths for the first contact in the center of indenter and for the complete contact. Dependently on the exponent of the power law, three cases are possible:

1. For $n<1$ no complete contact can be established regardless of how big the indentation depth is.

2. For $1<n<2$, the contact starts in the shape of very narrow ring and ins propagating inwards until the last contact is established in the center of the indenter.

3. For $n>2$, the contact starts as a thin ring which is then expanding inwards. However, at some indentation depth a contact is established in the center of the indenter. After that point, the central contact area expands outwards and the outer contact area propagates inwards until they merge.

This paper can also be considered as corrigendum to §2.5.16 of the book [5].